\documentclass[aps,prd,twocolumn,nofootinbib,superscriptaddress,letterpaper,preprintnumbers]{revtex4}
\pdfoutput=1
\usepackage[hypertex]{hyperref}
\usepackage{graphicx}

\begin{document}

\pagestyle{plain}

\preprint{UCB-PTH-08/09}

\title{The Quirky Collider Signals of Folded Supersymmetry}
\author{Gustavo Burdman}
\affiliation{Instituto de F\'{i}sica, Universidade de S\~{a}o Paulo,
R. do Mat\~{a}o 187, S\~{a}o Paulo, SP 05508-0900, Brazil}
\author{Z. Chacko}
\affiliation{Department of Physics, University of Maryland, College Park,
MD 20742}
\affiliation{Department of Physics, University of Arizona, Tucson, AZ 85721}
\author{Hock-Seng Goh}
\affiliation{Department of Physics, University of California,
Berkeley, and Theoretical Physics Group, LBNL, Berkeley, CA 94720}
\author{Roni Harnik}
\affiliation{
SLAC, Stanford University, Menlo Park, CA 94025}
\affiliation{SITP, Physics Department, Stanford University,
Stanford, CA 94305}
\author{Christopher A.~Krenke}
\affiliation{Department of Physics, University of Maryland, College Park,
MD 20742}
\affiliation{Department of Physics, University of Arizona, Tucson, AZ 85721}


\begin{abstract}

We investigate the collider signals associated with scalar quirks
(``squirks'') in folded supersymmetric models. As opposed to regular
superpartners in supersymmetric models these particles are uncolored, but
are instead charged under a new confining group, leading to radically
different collider signals.  Due to the new strong dynamics, squirks that
are pair produced do not hadronize separately, but rather form a highly
excited bound state. The excited ``squirkonium'' loses energy to radiation
before annihilating back into Standard Model particles. We calculate the
branching fractions into various channels for this process, which is
prompt on collider time-scales. The most promising annihilation channel
for discovery is W+photon which dominates for squirkonium near its ground
state. We demonstrate the feasibility of the LHC search, showing that the
mass peak is visible above the SM continuum background and estimate the
discovery reach.


\end{abstract}

 \maketitle


\section{Introduction}

One of the prime motivations for physics beyond the Standard Model (SM) at
scales accessible to the Large Hadron Collider (LHC) is the hierarchy
problem, namely, the quadratic sensitivity of the electroweak scale to
higher scales. The absence of any new physics up to very high scales would
imply an unnatural fine tuning of the Higgs potential. The hierarchy
problem has motivated a paradigm for model building whereby a new symmetry
is introduced at the TeV scale to protect the Higgs from quadratic
divergences. In such a scenario, naturalness is manifested in
cancellations between those SM loop diagrams which contribute to the Higgs
mass parameter and loops involving new physics. When considering the new
physics that the LHC may discover it suffices to consider only the
``little hierarchy'' between the scales set by the LHC reach and the
electroweak scale. For this purpose one loop cancellations are sufficient.

Within this paradigm the new states that cancel a given SM loop are
related to the corresponding SM particles by the symmetry. The largest
contribution to the Higgs mass parameter in the SM is from the top quark
loop, and therefore there will be new states related to the top by a
symmetry that cancel this loop. For example, if supersymmetry~\cite{susy}
is introduced the spectrum includes scalar top partners, $\tilde t$,
to cancel the top loop. 
Alternatively global symmetries,
or combinations of global and discrete symmetries, may be responsible for
protecting the Higgs, such as in little Higgs models~\cite{Little1,
Reviews} or
twin Higgs models~\cite{twin, LRtwin}. In these theories new fermionic
top partners, $t'$, are responsible for restoring naturalness.

In this class of theories the collider phenomenology crucially depends on
whether or not the top partners carry SM color. While in little Higgs
theories and supersymmetric theories the top partners are indeed charged
under SM color, in general this need not be the case. In particular, if
the symmetry that protects the Higgs mass involves a $Z_2$ symmetry
that interchanges SM color with a hidden color,
 \begin{equation} \label{z2}
{\rm SU(3)}_{\mathrm{QCD}} \leftrightarrow {\rm SU(3)}_{\mathrm{QCD'}}
 \end{equation}
where the symmetry acts on the SM SU(3) gauge fields and, more
generally, on all colored matter, the cancellation can take place
with top partners that are not charged under QCD, but only under
QCD$'$. The LHC signals are drastically different in such a
scenario since no new particle is strongly produced.

One possibility, realized in the mirror twin Higgs model~{\cite{twin}}, 
is that the top partners are not charged under any of the SM gauge 
groups. In this paper we focus on the alternative possibility, realized 
in folded supersymmetry~\cite{folded}, that the new quark partners are 
charged under QCD$'$, but retain SM SU(2$)_{\rm L}$ $\times$ U($)_{\rm 
Y}$ quantum numbers. The spectrum of states in this theory is similar to 
that of the MSSM, but with the crucial difference that the squarks are 
charged under SU(3$)_{\rm QCD'}$ instead of SM QCD. The typical scale for 
scalar quark masses is a few hundred GeV. Since the two QCD sectors are 
related by the symmetry of equation~(\ref{z2}) the scales $\Lambda$, 
where QCD gets strong, and ${\Lambda'}$, where QCD' gets strong, are 
close, with the difference arising from radiative effects. This implies 
there is a gap between the strong scale and the mass of the lightest 
matter field in the QCD$'$ sector. The absence of any light states 
charged under both the SM gauge groups and QCD$'$ implies this theory may 
be considered an example of a hidden valley model~\cite{hiddenvalley} in 
which the hidden valley particles are glueballs and in which the energy 
scales and particle spectrum are motivated directly by the hierarchy 
problem. The purpose of this paper is to study for the first time the 
collider phenomenology associated with the production of heavy quark 
partners in this scenario. The charge assignments and strong dynamics 
effects will lead to signals that are very different from either 
supersymmetric or generic hidden valley models.

Theories with a QCD$'$ sector where there is a large hierarchy
between the masses of the matter fields and the QCD$'$ scale,
$m_{q'}\gg\Lambda'$, give rise to very unusual
dynamics~\cite{bj}\cite{hiddenvalley}\cite{quirks}\cite{nussinov}. For this
reason the quarks (or scalar quarks) of such a sector have been
dubbed quirks (or squirks)~\cite{quirks}. To understand this, let
us first recall the dynamics of normal QCD. Consider two heavy
quarks that are produced back-to-back in a hard process.  As the
two quarks get farther apart and their distance approaches
$\Lambda^{-1}$, confining dynamics sets in and some of their
energy is lost to a gluonic flux tube extending between them. When
the local energy density in the flux tube is high enough it is
energetically favorable to pair create a light quark anti-quark
pair, ripping the tube. This mechanism of soft hadronization
allows the two heavy quarks to hadronize separately.

In quirky QCD, on the other hand, such a soft hadronization
mechanism is absent because there are no quarks with mass less
than or comparable to ~$\Lambda'$. The energy density in the QCD
flux tube, or more simply, the tension of the QCD string cannot
exceed $\Lambda'^2$ which is far less than the $m_{q'}$ per
Compton wavelength needed to create a heavy quirk anti-quirk pair.
The splitting of the QCD string by a quirk anti-quirk pair is
exponentially suppressed as $\exp
(-m_{q'}^2/\Lambda'^2)$~\cite{bj}. In fact, one may view the
entire process as single production of a highly excited bound
state, squirkonium. All of the kinetic energy that the quirks
posses at production, $\sqrt{\hat s}-2m_{q'}$, which is typically
of order $m_{q'}$, can be interpreted as squirkonium excitation
energy. This energy is radiated away into glueballs of QCD$'$ and
photons. Eventually the two quirks pair-annihilate back into
lighter states.

From the above discussion it is clear that the characteristic collider 
signatures of folded supersymmetry are determined by the final states 
that the squirks annihilate into.  In what follows we calculate the 
cross-section for production of these particles at the LHC, and evaluate 
the branching ratios for pair-annihilation into various final states. The 
possibility of discovering the soft photons from the loss of excitation 
energy will be discussed elsewhere~\cite{HW}. We then focus on the most 
promising annihilation channel for detection, which is W + photon, and 
demonstrate the reach for this search at the LHC. Some qualitative 
features of our analysis also apply to the supersymmetric model of Babu, 
Gogoladze and Kolda~{\cite{BGK}}, which also predicts quirky behaviour at 
the weak scale.

\section{Production and Energy Loss}

In a folded supersymmetric model, the scalar quirks have exactly
the same electroweak quantum numbers as the corresponding quarks,
but are charged under QCD$'$, not under QCD. Specifically, under
SU(3$)_{\rm C'}$ $\times$ SU(3$)_{\rm C}$ $\times$ SU(2$)_{\rm L}$
$\times$ U(1$)_{\rm Y}$, where SU(3$)_{\rm C'}$ corresponds to
QCD$'$, the quantum numbers of the squirks are
 \begin{eqnarray}
\tilde{Q} &&\; [3, 1, 2, (1/3)] \nonumber \\
\tilde{D}^c &&\; [\bar{3}, 1, 1, (2/3)] \nonumber \\
\tilde{U}^c \; && [\bar{3}, 1, 1, -(4/3)]
 \end{eqnarray}
Therefore, at a collider squirks are only weakly produced. This
could happen through the Drell-Yan process via an off-shell
photon, $Z$, or $W$, or alternatively by gauge boson fusion.
Production through the photon or the $Z$ typically does not result
in an observable signal, since the primary annihilation channel is
to glueballs of QCD$'$, which are invisible \footnote {Although
glueballs of QCD$'$ decay back to SM states, giving rise to
potentially observable signals~{\cite{hiddenvalley}}, in the
parameter range of interest this generally happens outside the
detector~{\cite{folded}}.}. 
In the simple analysis of this paper we focus on
production of squirks through s-channel $W^\pm$\footnote{It should be pointed out that weak boson fusion may dominate over Drell-Yan production at high squirk mass~\cite{VBF} and may require further study.}. In this case the
conservation of electric charge implies that squirk annihilation
must result in the emission of at least one charged particle.

\begin{figure}
\begin{center}
\includegraphics[
width=\columnwidth,
]{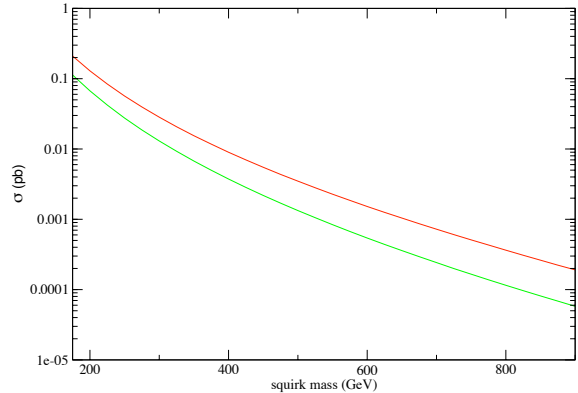}
\end{center}
\caption{ The total cross-section for production of first generation
squirk anti-squirk pairs via an s-channel $W^+$ (top curve) and
$W^-$ (bottom curve) at the LHC as a function
of the squirk mass. The up and down squirks have been taken to be
degenerate.}
\label{sigma}
\end{figure}

Figure~\ref{sigma} shows the production cross-section for first 
generation SU(2) doublet up-down squirk anti-squirk pairs through 
s-channel $W^+$ and $W^-$ as a function of the squirk mass. In folded 
supersymmetry the second generation squirks are expected to be nearly 
degenerate with those of the first which will effectively double the 
signal rate.  

The squirks are in general produced with kinetic energy much larger than
$\Lambda_{\rm QCD'}$. We expect that a significant fraction of this energy
is lost to photons and hidden glueballs prior to pair-annihilation. What
determines the time-scale on which this happens? In particular, should we
expect displaced vertices in quirky events?

We now argue that, in the context of folded supersymmetry, the
pair-annihilation of squirks is prompt on collider time-scales.
Assuming the excitations of squirkonium follow a Regge behavior
$E_n^2\sim n \Lambda'^2$, the energy spacing between adjacent
energy levels decreases like $1/\sqrt{n}$. In the limit of high
quantum number $n$ the energy levels may be treated as a continuum
and the system may be treated semi-classically. The semi classical
description is that of two massive charged particles connected by
a string with tension $\Lambda'^2$.  The string exerts a constant
force on the particles, leading to an acceleration of
$a=\Lambda'^2/m_{q'}$.

A system of accelerating charges must radiate soft photons as dictated
by Larmor's formula
 \begin{equation}
P=\frac{\partial E}{\partial t} = \frac{8 \pi \alpha}{3} a^2\,.
 \end{equation}
The time scale for energy loss from photon radiation, $E/P \sim m_{q'}/P$,
is thus
 \begin{equation}
t_{\mathrm rad}\sim \frac{3}{8\pi \alpha}\frac{m_{q'}^3}{\Lambda'^4}
=\left(\frac{m_{q'}}{\mbox{ 500 GeV}}\right)^3 \left(\frac{\mbox{ 5
GeV}}{\Lambda'}\right)^4 10^{-18}\,\mathrm{sec}
 \end{equation}
which is prompt. It is plausible that, because of the glueball
mass gap, radiation of soft photons contributes a substantial part
of the energy loss of excited squirkonium, leading to a novel
signature of this scenario~\cite{HW}. In any case, other
mechanisms of energy loss, such as emission of glueballs, can only
make $t_{\mathrm rad}$ shorter. The rate for pair-annihilation of
squirkonium from a low lying state is of order $m_{q'}
\alpha^3_{\rm QCD'} \alpha^2_{\rm W}$, which is also prompt on
collider time-scales. Here $\alpha_{\rm QCD'}$ and $\alpha_{\rm
W}$ are the structure constants for QCD$'$ and the SM weak
interactions respectively. We conclude that displaced vertices are
not a characteristic feature of folded supersymmetry.

\section{Annihilation}

Since squirkonium states produced through a W carry electric charge, we
expect that they will annihilate into SM particles, giving rise to
promising collider signals. However, we must first verify that the charged
state does not beta decay down to a neutral state prior to
re-annihilation, which would result in the absence of an observable
signal.

Since the first and second generation quarks have very small Yukawa
couplings, the mass splittings between the corresponding up-type and
down-type squirks in an SU(2) doublet are very small. The mass splitting is dominated by radiative
effects and is of order $ (e^2/16 \pi^2) (M_{\rm Z}^2/m_{q'})$. 
The rate for beta decay is
proportional to five powers of the mass splitting $\delta m_{q'}$, and is
therefore highly phase space suppressed.
 \begin{equation}
\Gamma = \frac{G_F^2 (\delta m_{q'})^5}{15\pi^3}
 \end{equation}
Therefore the scalar partners of the first two generations will
pair-annihilate before beta-decay can take place, resulting in an
observable signal. For the third generation squirks, on the other hand,
the mass splittings are larger, of order $m_t^2/(2m{q'})$ which is of order 30 GeV for squirks at 500 GeV.
The lifetime of a stop squirk with this mass splitting is of order $10^{-19}$ seconds, which may be comparable to the time scale for energy loss by radiation. The dynamics of a squirk pair which undergoes beta decay while oscillation may be quite interesting because of the large angular momentum introduced to the system, however we will not consider this possibility here. Even if the energy loss is much faster than beta decay, which may be the case because  glueball radiation, the stop-sbottom quirkonium state is likely to annihilate via a charged Higgs due to the large top yukawa coupling and is thus not likely to contribute to the signal we will consider bellow.
We therefore omit this case from further consideration,
and limit our discussion to the first two generations.

What are the possible final states? One possibility is that the squirk and
anti-squirk go back into two SM fermions through an s-channel W. In the
limit that the masses of the W and the final state fermions are neglected,
the total cross-section for a squirk anti-squirk pair into all possible SM
fermion-antifermion pairs is given by
 \begin{equation}
\sigma = N_{\mathrm{QCD'}}N_f v_{\rm rel} \; \frac{\pi \alpha_W^2 }{48 E^2}
 \end{equation}
Here $E$ is the energy of each squirk, $v_{\rm rel}$ is the relative
velocity between the squirks and $N_f = 12$ is the total number of
possible SM final states. Other possible annihilation channels include W +
photon and W + Z. Consider first the W + photon final state. In the limit
that $v_{\rm rel}$ is very small, we obtain
 \begin{equation}
\sigma v_{\rm rel} = N_{\mathrm{QCD'}}\frac{ \pi \alpha_W \alpha}{18 E^2}
 \end{equation}
In this expression we have neglected the mass of the W. Finally, for
the W + Z final state, in the limit of very small relative velocity, we
have
 \begin{equation}
\sigma v_{\rm rel} = N_{\mathrm{QCD'}}\,{\rm tan}^2 \theta_W \; \frac{ \pi
\alpha_W \alpha}{
18 E^2}
 \end{equation}
Here $\theta_W$ is the weak mixing angle, and we have again neglected the
masses of the W and the Z.  We see from these expressions that for
slow-moving squirks  which have
lost most of their energy or were produced close to threshold
the most important channel is to W + photon,
with annihilation to two fermions becoming significant only for larger
values of the relative velocity. 

\begin{figure}
\begin{center}
\includegraphics[width=\columnwidth]{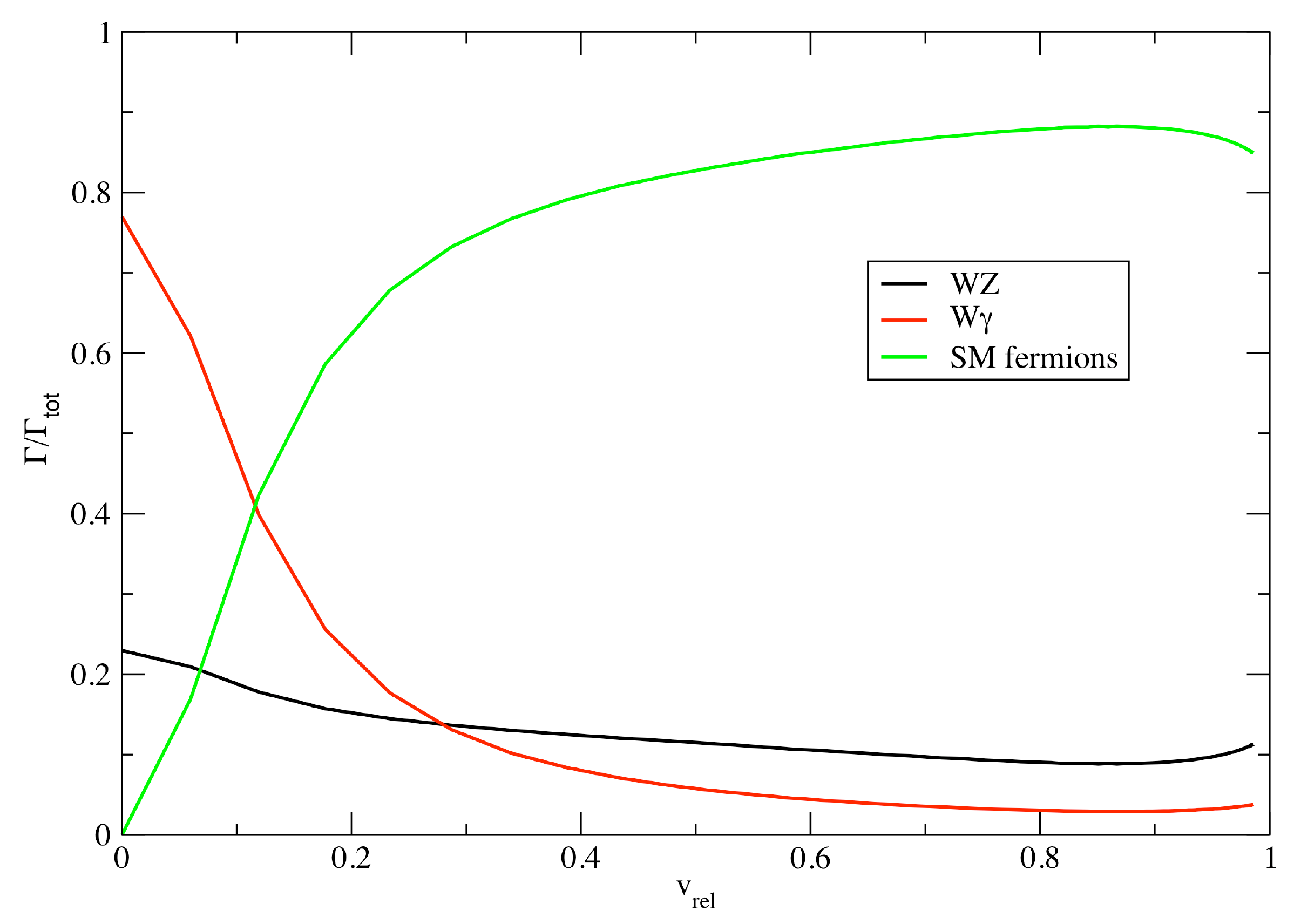}
\end{center}
\caption{Branching ratios for a charged squirk-antisquirk pair into
various final states, as a function of the relative velovity of the pair.}
\label{branching}
\end{figure}

Figure~\ref{branching} shows the branching ratios for pair-annihilation of
squirks into various final states, as a function of the relative velocity
of the incoming squirks. The relevant cross-sections have been evaluated
using CalcHEP~\cite{CALCHEP}. We see that for slow-moving squirks the
dominant annihilation channel is to W + photon, in agreement with our
analytic expressions above. This is promising for the LHC. The W + Z
channel is also large enough to play a role. For more energetic squirks
annihilation through an off-shell W into two SM fermions dominates. In
this case the charged lepton + missing energy and top + bottom final
states are most likely to prove useful. 

In order to devise a search strategy it is therefore important to understand whether annihilation typically occurs at high or low relative velocities.
As descirbed in more detail in~\cite{quirks, HW}, annihilation will occur dominantly at low values of angular momentum. The angular momentum in the squirkonium system will grow in a random walk as the excited state radiates quanta. Following the formalism developed in~\cite{quirks} we estimate that the probability for squirk  annihilation in a single period at low angular momentum and high velocity is approximately $10^{-4}$ \footnote{This is significantly less than the annihilation probability for fermionic quirks.}. This small probability implies that the squirk system will typically gain significant angular momentum before annihilation occurs, and the squirkonium pair will reach near its ground state before annihilating. With this motivation in mind we will consider a search for squirks in the W+photon annihilation channel in the next section.

Before moving on to collider searches we should briefly comment on the 
possibility of annihilating to W+glueball. The direct annihilation to 
W+gluon naively vanishes because any such amplitude is proportional to 
the trace of a single SU(3) generator. Annihilating to W+glueball thus 
requires emitting at least two gluons, and is therefore phase space 
suppressed. However, there may be another possibility for this 
annihilation to occur. It is believed that hybrid states exist in 
QCD~\cite{PDGhybrid}. These exotic states may be viewed as bound states 
consisting of a meson in the color octet state and a gluon. Lattice 
studies show that QCD hybrid states carry a variety of spin and parity 
quantum numbers, including some states that mix directly with 
conventional mesons, and some that do not. In regular QCD the states that 
do not mix with mesons can decay rapidly to, for example, pions. However 
in quirky QCD$'$, these states may be semi-stable. In particular, lattice 
studies show that the mass splitting between some hybrid states and their 
corresponding (same flavor) conventional meson state is smaller than the 
estimated QCD glueball mass~\cite{PDGhybrid}. If this is indeed the case, 
some quirky hybrid state will be stable to glueball emission due to 
kinematics and will thus annihilate directly. Since the $q' q'^*$ state 
is in the color octet, the amplitude for single gluon emission is no 
longer a trace of a single generator and will not vanish. These states 
are thus most likely to annihilate to W+glueball. The dynamical question 
is then, how often does the radiative energy loss of excited squirkonium 
lead to one of the exotic hybrid state rather than the conventional 
ground state? Keeping the W+glueball channel in mind we leave this 
question for further work.

\section{LHC Search}

In this final section we will demonstrate the feasibility of searching for
the annihilation products of squirk bound states at the LHC and estimate
the signal to background ratios after employing various kinematic cuts. We
will focus on the W+photon annihilation signal. As discussed above, this
final state dominates when the squirkonium bound state has lost most of
its energy by radiation and the annihilation takes place at or near the
ground state. The expected signal is a \emph{peak} in the invariant
mass of W+photon. The concentration of the signal at a mass peak will be
crucial for the signal to stand above background.

We focus on events in which the W boson decays leptonically. Despite the 
inability to measure the neutrino energy and longitudinal momentum the 
invariant mass of the squirkonium may be reconstructed if one 
\emph{assumes} the lepton and missing $E_T$ come from an on shell W, and 
that the neutrino is the only source of missing transverse energy. If 
this is the case, the width of the mass peak arises predominantly form 
detector effects and from squirkonium decays from 
low lying excited states, rather than true ground states (of order 
$\Lambda'\sim$ 10 GeV).

This mass peak is likely to be smeared due to additional transverse
missing energy from the radiative decay. The energy loss due to radiation
is expected to be distributed in a particular pattern which is symmetric
under $\vec x \to -\vec x$, that is, for every glueball or photon emitted
in one direction, a similar photon will be emitted in the opposite
direction. In the classical limit (an infinite number of soft quanta) such
energy loss does not introduce new transverse momentum. However, for
glueball radiation the classical limit is not appropriate because the
glueball mass cannot be neglected when compared to the total amount of
energy radiated. The additional missing transverse energy is inversely
proportional to the square root of the number of glueball quanta emittted,
in the limit that the total energy radiated is held fixed. We will
(conservatively) assume that most of the energy is lost by glueball
emission giving a missing $E_T$ of order $\sqrt{m_{q'}\Lambda'}$, which is
of order 100 GeV or less in most of our parameter space.

\begin{figure}
\begin{center}
\includegraphics[width=\columnwidth]{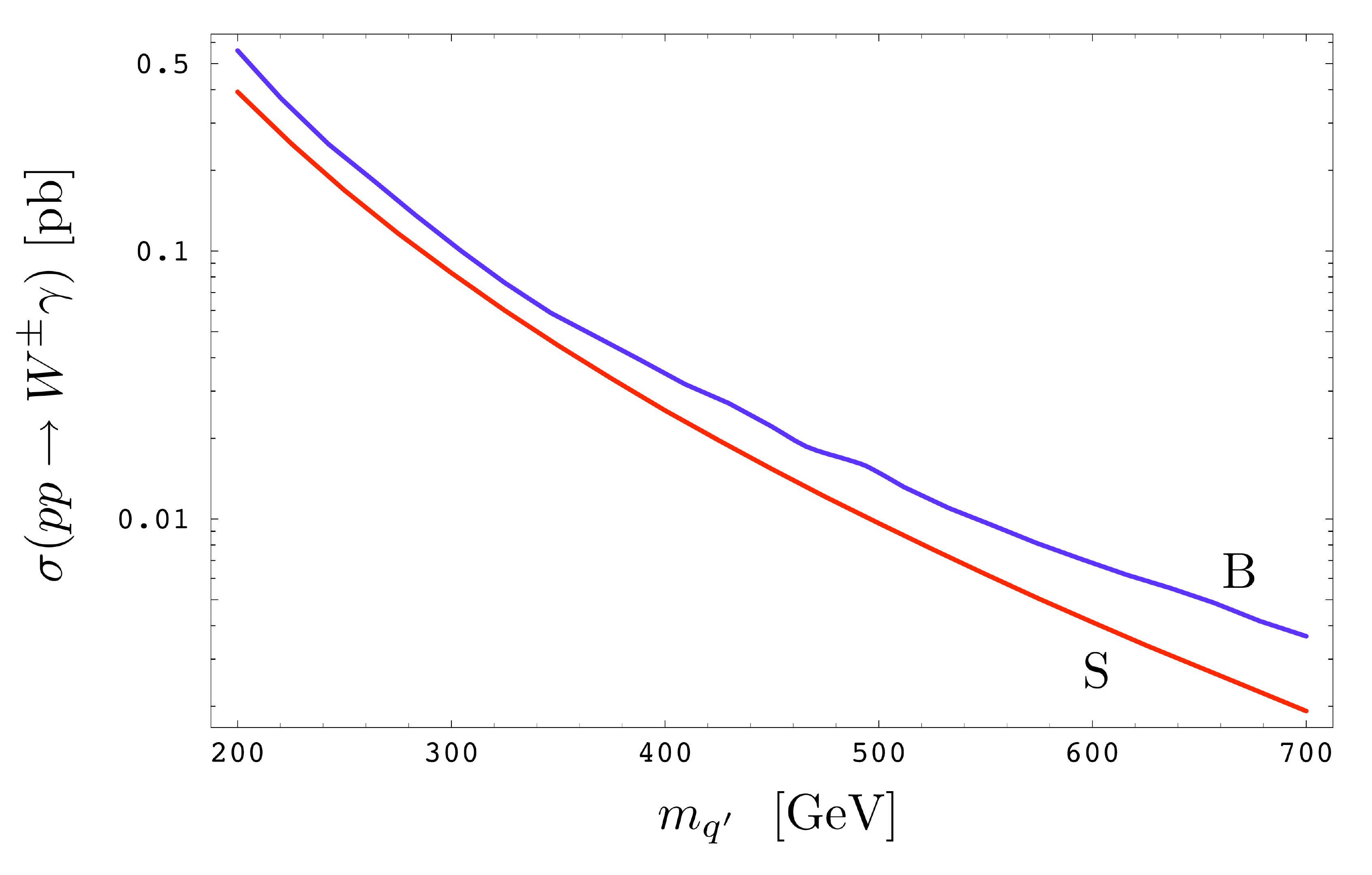}
\end{center}
\caption{
An estimate of the signal versus the background as a function of the
squirk mass in GeV. The bottom curve is the squirk pair production cross section
in the charged channel. The annihilation of these squirk pair will
dominantly produce $W^\pm$+photon with and invariant mass of $\sim
2m_{q'}$. The top curve is the SM $W^\pm$+photon with
$|m_{W\gamma}-2m_{q'}|< \sqrt{\Lambda m_{q'}}$ for $\Lambda=15$ GeV.}
\label{SIGvsBG}
 \end{figure}

In the case where the W decays leptonically, the leading background is
continuum SM production of W+photon. For simplicity we will
estimate the signal-to-background ratio by comparing the W+photon signal
cross section to the invariant mass distributions of background W+photon
in the mass window a $|m_{W\gamma}-2m_{q'}|<
\sqrt{\Lambda m_{q'}} \mbox{ GeV}$. For the purpose of this estimate have conservatively chosen a rather high value of $\Lambda=15 GeV$, about a factor of 10 above our QCD gluball mass calculated on the lattice.
Background events were generated using
MadEvent~\cite{ME}. As shown in Figure~\ref{SIGvsBG}, the signal to
background ratio is of order a half for all of the mass range within LHC
reach. 
Taking the leptonic branching fraction of the W boson (to electrons or
muons only) and a low branching fraction of squirkonium to W+photon of
0.6, a 5 sigma discovery of squirks with a mass of 400 GeV is possible
with $\sim$11 fb$^{-1}$ of data. With 100 fb$^{-1}$ the discovery reach is
approximately 620 GeV.

\begin{figure}
\begin{center}
\includegraphics[width=\columnwidth]{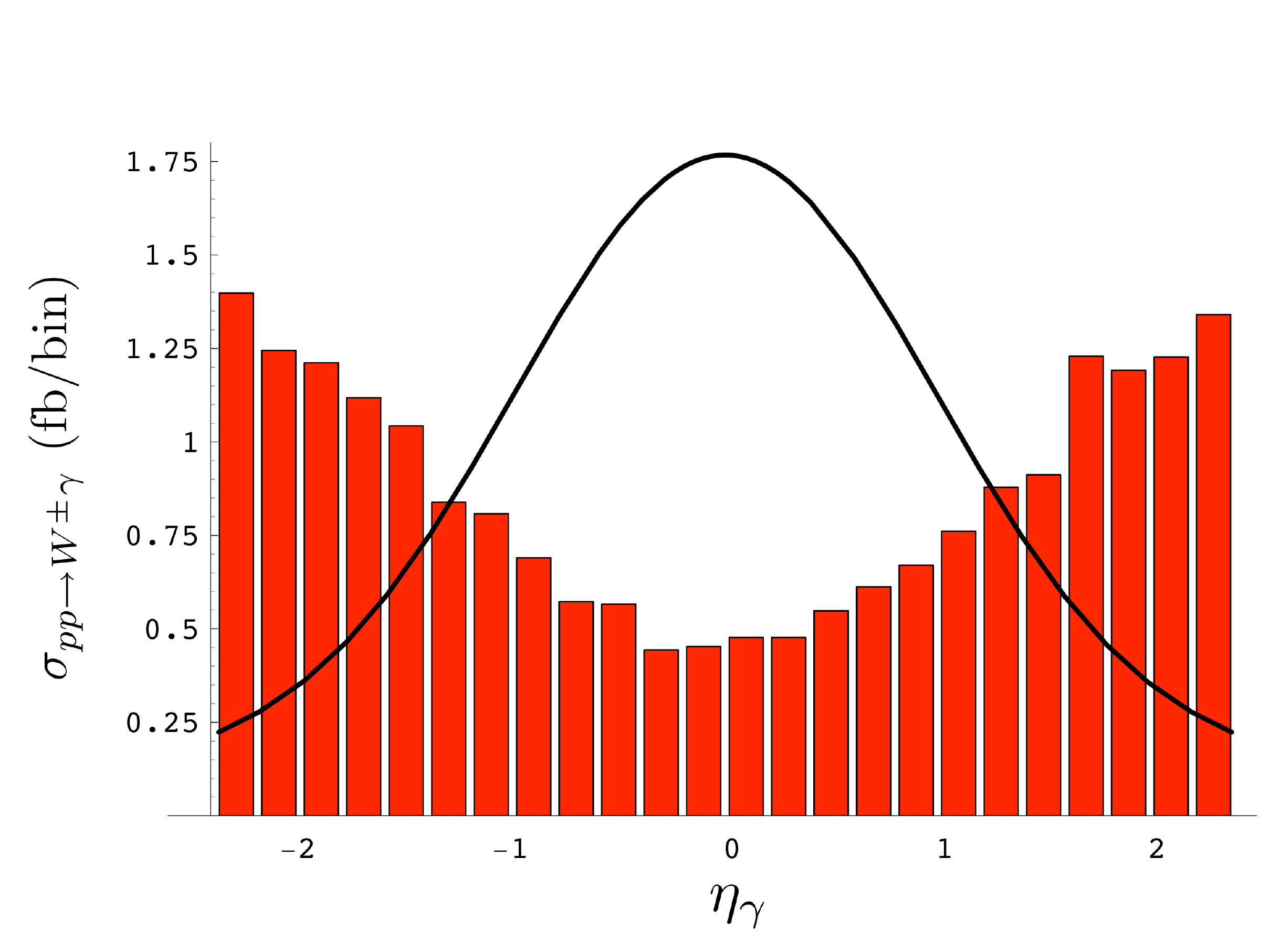}
\end{center}
\caption{
The photon $\eta$ distribution for SM W+photon events with center of mass
energy of 1 TeV or more (red bars). The estimated signal distribution is
shown as a solid curve. Employing a cut of $\eta<1.5$ increases the signal
to background ration by more than a factor of two.}
\label{etaplot}
\end{figure}

One can improve the discovery reach by employing a pseudorapidity cut on
the final state photon. The signal events involve the production and decay
of heavy squirkonium. Even after losing its excitation energy, the
squirkonium will be nearly at rest in the lab frame and is expected to
decay isotropically.  In Figure~\ref{etaplot} we show the distribution of
the photon's pseudorapidity, $\eta$, for SM W+photon events with invariant
mass of 1 TeV or more. The shape of the distribution is similar for
different invariant masses. This distribution is compared with the
expected signal distribution (shown as solid curve). This was estimated by
convoluting the $\eta$ distribution of isotropic decay with the
longitudinal boost of the squirkonuim system (which is simply the
longitudinal boost of the CoM frame in squirk production, generated by
MadEvent).

Placing a cut on $\eta$ of the photon of 1.5 reduces the background by a
factor of $\sim2.2$ while the signal is only reduced by $\sim 15\%$.
Including the $\eta$ cut, 5 sigma discovery of 400 GeV squirks may be
reached with approximately 8.5 fb$^{-1}$. Taking this estimate at face
value, with 100 fb$^{-1}$ the LHC will be able to discover one generation
of squirks up to a mass of 500 GeV and two degenerate generations of
squirks (as is the case in in folded SUSY) up to a mass of 650 GeV.
However, one could imagine improving this analysis, for example, by
optimization of the $\eta$ cut. Alternatively, we could perform additional
cuts on lepton rapidity or compare transverse mass distributions instead
of reconstructed invariant masses~\cite{Tmass}. Observation of soft
photons from the radiative decay of squirkonium could also reduce the
background and thereby increase the reach~\cite{HW}.


In summary, we have performed the first detailed study of the collider
phenomenology associated with squirks in folded supersymmetric models, and
identified several promising discovery channels for the LHC.
\\[10pt]
{\bf Acknowledgments} 
It is a pleasure to thank Johan Alwall, Elliott Cheu, Markus Luty, Shmuel Nussinov and 
Michael Peskin for useful discussions. GB acknowledges the support of the 
State of S\~{a}o Paulo Research Foundation (FAPESP), as well as the 
Brazilian National Counsel for Technological and Scientific Development 
(CNPq). The work of ZC and CAK was partially supported by the NSF under 
award number PHY-0408954. The work of HSG was supported in part by the 
NSF grant PHY-04-57315 and by DOE under contract DE-AC02-05CH11231. The 
work of RH was supported by DOE grant DE-AC02-76SF00515. RH and HSG also wish to thank the Aspen Center for physics where some of this work was conducted.

\end{document}